# Manifestation of the Upper Hubbard band in the 2D Hubbard model at low electron density (review-article)

M.Yu. Kagan[1], V.V. Val'kov[2], P. Woelfle[3]

[1]P.L. Kapitza Institute for Physical Problems, Moscow, Russia

[2]Kirenskii Institute of Physics, Krasnoyarsk, Russia

[3]Institute for Theory of Condensed Matter and Center for Functional Nanostructures, Karlsruhe Institute of Technology, Karlsruhe, Germany

**Abstract.**

We consider the 2D Hubbard model in the strong-coupling case ($U \gg W$) and at low electron density ($nd^2 \ll 1$). We find an antibound state as a pole in the two-particle T-matrix. The contribution of this pole in the self-energy reproduces a two-pole structure in the dressed one-particle Green-function similar to the Hubbard-I approximation. We also discuss briefly the Engelbrecht-Randeria mode which corresponds to the pairing of two holes below the bottom of the band for $U \gg W$ and low electron density. Both poles produce non-trivial corrections to Landau Fermi-liquid picture already at low electron density but do not destroy it in 2D.

**Introduction.**

At low electron density ($nd^2 \ll 1$ - practically empty band) and in the strong-coupling case $U \gg W$ the effective interactions in the 2D Hubbard model [1] can be described in the T-matrix approximation (see Kanamori [2]). In the low energy sector $\varepsilon \leq \varepsilon_F$ and in the framework of this description the 2D Hubbard model becomes equivalent to a 2D Fermi-gas with quadratic spectrum and short-range repulsion [3]. Thus it can be characterized by the 2D gas-parameter of Bloom [4]:

$$f_0 \approx \frac{1}{\ln\frac{1}{nd^2}} \; ; \qquad (1)$$

where $n = \frac{p_F^2}{2\pi}$ is the electron density in 2D (for both spin projections, taking into account that $n_\sigma = n_{-\sigma} = \frac{n}{2}$ in the unpolarized case), $p_F$ is the Fermi-momentum, $d$ - is the intersite distance. Accordingly many properties of the 2D Hubbard model at low electron density, and in particular the quasiparticle damping near the Fermi-surface $\gamma \sim \mathrm{Im}\Sigma(\varepsilon_p, \vec{p}) \sim f_0^2 \frac{\varepsilon_p^2}{\varepsilon_F} \ln\left|\frac{\varepsilon_F}{\varepsilon_p}\right|$ have Landau Fermi-liquid character (amended with the specific 2D logarithm) [5], where $\varepsilon_p = \frac{p^2}{2m} - \varepsilon_F$ is quasi-particle spectrum in the low-energy sector $\varepsilon \leq \varepsilon_F$ and $f_0$ is given by (1). Correspondingly the averaging of $\mathrm{Im}\Sigma(\varepsilon_p, \vec{p})$ with Fermionic distribution function $n_F(\frac{\varepsilon_p}{T})$ produces the familiar result $\gamma(T) \sim \mathrm{Im}\Sigma(T) \sim T^2 \ln T$ in 2D. Accordingly the quasiparticle residue $Z \sim \left(1 - \frac{\partial \mathrm{Re}\Sigma}{\partial \omega}\right)^{-1}$ is nonvanishing for $\omega \to 0$. However, as first mentioned by J. Hubbard [1] and P.W. Anderson [6], for $U \gg W$ the presence of a band of a finite width produces at high energies an additional pole in the two-particle T-matrix, well separated from all other poles, with the energy:

$$\varepsilon \sim U > 0. \qquad (2)$$

This pole is usually called the antibound state. Already in the first iteration of the self-consistent T-matrix approximation this pole yields a non-trivial contribution to the self-energy $\Sigma(\varepsilon, \vec{p})$. As a result the dressed one-particle Green-function acquires a two-pole structure, very similar to the Hubbard-I approximation [1].

**The Theoretical model.**

We consider the simplest 2D Hubbard model on the square lattice:

$$\hat{H}' = \hat{H} - \mu \hat{N} = -t \sum_{\langle ij \rangle \sigma} c_{i\sigma}^+ c_{j\sigma} + U \sum_i n_{i\uparrow} n_{i\downarrow} - \mu \sum_{i\sigma} n_{i\sigma}, \qquad (3)$$

where $n_{i\sigma} = c_{i\sigma}^+ c_{i\sigma}$ - is the density operator of electrons on site $i$ with spin-projection $\sigma$, $U$ - is Hubbard repulsion, $t$ - is hopping integral, $\mu$ - is the chemical potential. The bandwidth $W = 8t$ on the square lattice. After Fourier-transforming we get:

$$\hat{H}' = \sum_{p\sigma} \varepsilon_p c_{p\sigma}^+ c_{p\sigma} + U \sum_{pp'q} c_{p\uparrow}^+ c_{p'\downarrow}^+ c_{p-q\downarrow} c_{p'+q\uparrow}, \qquad (4)$$

where $\varepsilon_p = -2t(\cos p_x d + \cos p_y d) - \mu$ - is the quasiparticle spectrum of the uncorrelated problem. For low electron density $p_F d \ll 1$ we can often use the quadratic approximation for the spectrum:

$$\varepsilon_p = \frac{p^2 - p_F^2}{2m}, \qquad (5)$$

where $m = \dfrac{1}{2td^2}$ is the band-mass; $\mu = -\dfrac{W}{2} + \varepsilon_F$ is chemical potential and $t_p = \varepsilon_p + \mu = -2t(\cos p_x d + \cos p_y d) \approx -\dfrac{W}{2} + tp^2 d^2 = -\dfrac{W}{2} + \dfrac{p^2}{2m}$. We will mostly consider the physically more transparent strong-coupling case $U \gg W$ at low electron density $nd^2 \ll 1$.

**T-matrix approximation.**

We start with the standard definition of the T-matrix in 2D[4,7]:

$$T = \frac{Ud^2}{1 - Ud^2 \int \dfrac{d^2 \vec{p}}{(2\pi)^2} \dfrac{1 - n_{F\sigma}(\varepsilon_p) - n_{F-\sigma}(\varepsilon_{q-p})}{(\omega - \varepsilon_p - \varepsilon_{q-p} + io)}}, \qquad (6)$$

The poles of the T-matrix are governed by the condition:

$$1 = Ud^2 \int \frac{d^2 \vec{p}}{(2\pi)^2} \frac{1 - n_{F\sigma}(\varepsilon_p) - n_{F-\sigma}(\varepsilon_{q-p})}{(\omega - \varepsilon_p - \varepsilon_{q-p} + io)}. \qquad (7)$$

For the antibound state for which $\omega \sim U$ we can expand (7) and get (see also [8]):

$$1 = Ud^2 \int \frac{d^2 \vec{p}}{(2\pi)^2} \frac{1 - n_{F\sigma}(\varepsilon_p) - n_{F-\sigma}(\varepsilon_{p-q})}{\omega} \left[ 1 + \frac{t_p + t_{p-q} - 2\mu}{\omega} \right], \qquad (8)$$

where $\varepsilon_p = t_p - \mu$; $\varepsilon_{p-q} = t_{p-q} - \mu = \varepsilon_{q-p}$.

Equivalently we can write:

$$1 = \frac{Ud^2}{\omega} \int \frac{d^2\vec{p}}{(2\pi)^2} \Big[1 - n_{F\sigma}(\varepsilon_p) - n_{F-\sigma}(\varepsilon_{p-q})\Big] +$$
$$+ \frac{Ud^2}{\omega^2} \int \frac{d^2\vec{p}}{(2\pi)^2} \Big[1 - n_{F\sigma}(\varepsilon_p) - n_{F-\sigma}(\varepsilon_{p-q})\Big](t_p + t_{p-q} - 2\mu) \quad (9)$$

and use that $\int \frac{d^2\vec{p}}{(2\pi)^2}(t_p + t_{p-q}) = 0$ when we integrate over the Brillouin zone. Thus:

$$1 = \frac{Ud^2}{\omega d^2}(1 - \frac{nd^2}{2} - \frac{nd^2}{2}) + \frac{Ud^2}{\omega^2} \int \frac{d^2\vec{p}}{(2\pi)^2}(-2\mu) -$$
$$- \frac{Ud^2}{\omega^2} \int \frac{d^2\vec{p}}{(2\pi)^2} \Big[n_{F\sigma}(\varepsilon_p) + n_{F-\sigma}(\varepsilon_{p-q})\Big]\Big[\varepsilon_p + \varepsilon_{p-q}\Big] \quad (10)$$

where we used that in unpolarized case $n_\sigma = n_{-\sigma} = \frac{n}{2}$. Note that

$$1 = d^2 \int_{BZ} \frac{d^2\vec{p}}{(2\pi)^2} = d^2 \int_{-\pi/d}^{\pi/d} \frac{dp_x}{2\pi} \int_{-\pi/d}^{\pi/d} \frac{dp_y}{2\pi} \text{ - for the integration over the Brillouine zone. Hence:}$$

$$1 = \frac{U}{\omega}(1 - nd^2) - \frac{U2\mu}{\omega^2} - \frac{Ud^2}{\omega^2} \int \frac{d^2\vec{p}}{(2\pi)^2}\Big[n_{F\sigma}(\varepsilon_p) + n_{F-\sigma}(\varepsilon_{p-q})\Big]\Big[\varepsilon_p + \varepsilon_{p-q}\Big]. \quad (11)$$

In the third term of (11) the integration is restricted by Fermi-factors and hence we can use quadratic approximation for the spectrum $\varepsilon_p = \frac{p^2}{2m} - \varepsilon_F$. Then for the third term we get:

$$-\frac{Ud^2}{\omega^2} N_{2D}(0)[2\int_{-\varepsilon_F}^{0} \varepsilon_p d\varepsilon_p + \int_{-\varepsilon_F}^{0} d\varepsilon_p \int_0^{\pi} \frac{d\varphi}{\pi}\Big(\frac{p^2 + q^2 - 2pq\cos\varphi}{2m} - \varepsilon_F\Big) +$$
$$+ \int_{-\varepsilon_F}^{0} d\varepsilon_p \int_0^{\pi} \frac{d\varphi}{\pi}\Big(\frac{p^2 + q^2 + 2pq\cos\varphi}{2m} - \varepsilon_F\Big)] = -\frac{Ud^2}{\omega^2} N_{2D}(0)[4\int_{-\varepsilon_F}^{0} \varepsilon_p d\varepsilon_p + 2\int_{-\varepsilon_F}^{0} d\varepsilon_p \frac{q^2}{2m}] = \quad (12)$$
$$= -\frac{Ud^2}{\omega^2} N_{2D}(0)[-2\varepsilon_F^2 + 2\varepsilon_F \frac{q^2}{2m}]$$

where we used that $\int n_{F-\sigma}(\varepsilon_{p-q})\varepsilon_p \frac{d^2\vec{p}}{(2\pi)^2} = \int n_{F-\sigma}(\varepsilon_p)\varepsilon_{p+q} \frac{d^2\vec{p}}{(2\pi)^2}$.

In (12) $N_{2D}(0) = \frac{m}{2\pi}$ - is the density of states in 2D for the quadratic spectrum. Hence:

$$1 = \frac{U}{\omega}(1 - nd^2) - \frac{U 2\mu}{\omega^2} - \frac{Ud^2}{\omega^2}\frac{m\varepsilon_F}{2\pi}[-2\varepsilon_F + \frac{q^2}{m}]. \tag{13}$$

Having in mind that $\frac{m\varepsilon_F}{2\pi} = \frac{p_F^2}{4\pi} = \frac{n}{2}$ we get

$$1 = \frac{U}{\omega}(1 - nd^2) - \frac{U 2\mu}{\omega^2} - \frac{Und^2}{2\omega^2}\left(-2\varepsilon_F + \frac{q^2}{m}\right) \tag{14}$$

Accordingly for the antibound state:

$$\omega_{ab} \approx U(1 - nd^2) - \frac{U 2\mu}{U(1 - nd^2)} + \frac{Und^2}{2U(1 - nd^2)}\left(2\varepsilon_F - \frac{q^2}{m}\right) \tag{15}$$

Or respectively

$$\omega_{ab} \approx U(1 - nd^2) - \frac{2\mu}{(1 - nd^2)} + \frac{nd^2}{(1 - nd^2)}\left(\varepsilon_F - \frac{q^2}{2m}\right) =$$

$$= U(1 - nd^2) - 2\mu - \frac{nd^2 2\mu}{(1 - nd^2)} + \frac{nd^2}{(1 - nd^2)}\left(\varepsilon_F - \frac{q^2}{2m}\right) = \tag{16}$$

$$= U(1 - nd^2) - 2\mu + \frac{nd^2}{(1 - nd^2)}(\varepsilon_F - 2\mu) - \frac{nd^2}{(1 - nd^2)}\frac{q^2}{2m}$$

By analogy with attractive-U Hubbard model [9] we can introduce "bosonic" chemical potential:

$$\mu_B = 2\mu - |E_b|, \tag{17}$$

where $|E_b| = U(1-nd^2) + \dfrac{nd^2}{(1-nd^2)}(\varepsilon_F - 2\mu) \approx U(1-nd^2) + nd^2 W$ (18)

is a "binding" energy of antibound pair and $-\dfrac{q^2}{4m^*}$ for the spectrum, where the effective mass reads:

$$m^* = m\dfrac{(1-nd^2)}{2nd^2} \gg m \text{ for } nd^2 \ll 1 \quad (19)$$

Then we can represent:

$$\omega_{ab} = |E_b| - 2\mu - \dfrac{q^2}{4m^*} = -\dfrac{q^2}{4m^*} - \mu_B, \quad (20)$$

which is quite nice. The spectrum (20) closely resembles the pole of the attractive-U Hubbard model for $|E_b| > \varepsilon_F$ [9]. The important difference is, however, in the relative sign between $2\mu$ and $|E_b|$. In the attractive-U Hubbard model $\mu_B = 2\mu + |E_b|$ and the real pairs are created below the bottom of the band. Thus $\mu \approx -\dfrac{|E_b|}{2}$ and $\mu_B \to 0$ at low temperatures. In the repulsive-U Hubbard model for low electron density $nd^2 \ll 1$: $\mu \approx -\dfrac{W}{2} + \varepsilon_F$ for low temperatures. Only in the case of half-filled band $nd^2 = 1$ (one electron per site) the chemical potential $\mu \approx U/2$ "jumps" in the middle of the Mott-Hubbard gap $\Delta_{MH} = U$. The situation resembles that for a semiconductor: the chemical potential for $nd^2 = 1$ lies in the middle of the forbidden gap. Another important difference is connected with the hole-like dispersion in (20) that is with the sign "-" in front of $\dfrac{q^2}{4m^*}$.

The T matrix close to the pole reads [9]:

$$T(\omega, \vec{q}) \approx \dfrac{U\omega}{\left(\omega + \dfrac{q^2}{4m^*} + \mu_B + io\right)}. \quad (21)$$

**Imaginary part of the self-energy.**

In the first iteration to the self-consistent T-matrix approximation (see [10,11]):

$$\text{Im}\Sigma(\omega,\vec{k}) = \int \frac{d^2\vec{p}}{(2\pi)^2} \text{Im}T(\omega+\varepsilon_p, \vec{p}+\vec{k})\left[n_F(\varepsilon_p) + n_B(\varepsilon_p+\omega)\right] =$$

$$= \pi \int \frac{d^2\vec{p}}{(2\pi)^2} U(\omega+\varepsilon_p)\delta(\omega+\varepsilon_p+\mu_B+\frac{(\vec{p}+\vec{k})^2}{4m^*})\left[n_F(\varepsilon_p) + n_B\left(-\frac{(\vec{p}+\vec{k})^2}{4m^*}-\mu_B\right)\right]$$ (22)

where $n_F(\varepsilon_p)$ - is fermionic distribution function, $n_B\left(-\frac{(\vec{p}+\vec{k})^2}{4m^*}-\mu_B\right)$ - is bosonic distribution function. Having in mind that $\mu_B = 2\mu - |E_b| \sim -U$ we get for $U >> T$ [11]:

$$n_B\left(-\frac{(\vec{p}+\vec{k})^2}{4m^*}-\mu_B\right) = \frac{1}{\left(e^{-\frac{(\vec{p}+\vec{k})^2}{4m^*T}} e^{-\frac{\mu_B}{T}} - 1\right)} \to 0.$$ Thus:

$$\text{Im}\Sigma(\omega,\vec{k}) = \pi U \int \frac{d^2\vec{p}}{(2\pi)^2}\left[-\frac{(\vec{p}+\vec{k})^2}{4m^*}-\mu_B\right]\delta(\omega+\varepsilon_p+\mu_B+\frac{(\vec{p}+\vec{k})^2}{4m^*})n_F(\varepsilon_p)$$

(23)

Here again we have the important difference with attractive-U Hubbard model where at low temperatures $T \to 0$: $2n_B = n$ while $n_F = 0$. In repulsive-U Hubbard model we have vice-versa $n_B = 0$ and $n = n_F$ for $T \to 0$.

Having in mind that $m^*/m >> 1$ for $nd^2 << 1$ we can neglect $\frac{(\vec{p}+\vec{k})^2}{4m^*}$ in (23). Thus we get [11]:

$$\text{Im}\Sigma(\omega,\vec{k}) = -\pi N_{2D}(0)U\mu_B \int_{-\varepsilon_F}^{0} d\varepsilon_p \delta(\omega+\varepsilon_p+\mu_B) =$$

$$= -\pi N_{2D}(0)U\mu_B\left[\theta(\omega+\mu_B) - \theta(\omega+\mu_B-\varepsilon_F)\right]$$ (24)

**Real part of the self energy.**

Correspondingly for the real part of the self-energy[10,11]:

$$\text{Re}\Sigma(\omega,\vec{k}) = \int \text{Re}T(\omega+\varepsilon_p, \vec{p}+\vec{k})\left[n_B(\omega+\varepsilon_p) + n_F(\varepsilon_p)\right]\frac{d^2\vec{p}}{(2\pi)^2} \quad (25)$$

and again neglecting $n_B(\omega+\varepsilon_p)$ for $U >> T$ we get:

$$\text{Re}\Sigma(\omega,\vec{k}) = UN_{2D}(0)\int n_F(\varepsilon_p)d\varepsilon_p \frac{(\omega+\varepsilon_p)}{\left(\omega+\varepsilon_p+\mu_B+\frac{(\vec{p}+\vec{k})^2}{4m^*}\right)}. \quad (26)$$

For $m^*/m >> 1$ : $\frac{(\vec{p}+\vec{k})^2}{4m^*}$ is small and thus:

$$\text{Re}\Sigma(\omega,\vec{k}) = UN_{2D}(0)\int_{-\varepsilon_F}^{0} d\varepsilon_p \frac{(\omega+\varepsilon_p)}{(\omega+\varepsilon_p+\mu_B)} = UN_{2D}(0)\left[\varepsilon_F - \mu_B\int_{-\varepsilon_F}^{0}\frac{d\varepsilon_p}{(\omega+\varepsilon_p+\mu_B)}\right] =$$
$$= UN_{2D}(0)\left[\varepsilon_F - \mu_B\ln\frac{|\omega+\mu_B|}{|\omega+\mu_B-\varepsilon_F|}\right] \quad (27)$$

Assuming that $|\omega+\mu_B| > \varepsilon_F$ and expanding the logarithm in the second term we get:

$$\text{Re}\Sigma(\omega,\vec{k}) = UN_{2D}(0)\frac{\varepsilon_F\omega}{\omega+\mu_B} = U\frac{nd^2}{2}\frac{\omega}{\omega+\mu_B} \quad (28)$$

Thus the pole of the dressed one-particle Green-function [12] $G^{-1}(\omega,\vec{k}) = G_0^{-1}(\omega,\vec{k}) - \Sigma(\omega,\vec{k})$ reads:

$$\omega - \varepsilon_k - U\frac{nd^2}{2}\frac{\omega}{\omega+\mu_B} = 0 \quad (29)$$

Correspondingly:

$$\omega^2 + \left(\mu_B - \varepsilon_k - \frac{Und^2}{2}\right)\omega - \varepsilon_k \mu_B = 0$$

$$\left(\omega + \frac{\mu_B - \varepsilon_k - \frac{Und^2}{2}}{2}\right)^2 - \left(\frac{\mu_B - \varepsilon_k - \frac{Und^2}{2}}{2}\right)^2 - \varepsilon_k \mu_B = 0$$

As a result:

$$\omega_{1,2} = -\frac{\left(\mu_B - \varepsilon_k - \frac{Und^2}{2}\right)}{2} \pm \sqrt{\left(\frac{\mu_B - \varepsilon_k - \frac{Und^2}{2}}{2}\right)^2 + \varepsilon_k \mu_B} \quad (30)$$

Having in mind that $\mu_B \sim -U$ we can expand the square root in (30). Then:

$$\omega_{1,2} = -\frac{\left(\mu_B - \varepsilon_k - \frac{Und^2}{2}\right)}{2} \pm \left(\left|\frac{\mu_B - \varepsilon_k - \frac{Und^2}{2}}{2}\right| + \frac{\varepsilon_k \mu_B}{\left|\mu_B - \varepsilon_k - \frac{Und^2}{2}\right|}\right) \quad (31)$$

We know that $\mu_B < 0$ and $|\mu_B| \gg \left\{|\varepsilon_k|; \frac{Und^2}{2}\right\}$.

That is why $\left|\frac{\mu_B - \varepsilon_k - \frac{Und^2}{2}}{2}\right| = -\frac{\left(\mu_B - \varepsilon_k - \frac{Und^2}{2}\right)}{2}$ and hence:

$$\omega_{1,2} = -\frac{\left(\mu_B - \varepsilon_k - \frac{Und^2}{2}\right)}{2} \mp \left(\frac{\mu_B - \varepsilon_k - \frac{Und^2}{2}}{2} + \frac{\varepsilon_k \mu_B}{\mu_B - \varepsilon_k - \frac{Und^2}{2}}\right) \quad (32)$$

Finally:

$$\begin{cases} \omega_1 = -\left(\mu_B - \varepsilon_k - \dfrac{Und^2}{2}\right) - \dfrac{\varepsilon_k \mu_B}{\mu_B - \varepsilon_k - \dfrac{Und^2}{2}} \\ \omega_2 = \dfrac{\varepsilon_k \mu_B}{\mu_B - \varepsilon_k - \dfrac{Und^2}{2}} \end{cases} \quad (33)$$

The dressed Green-function $G(\omega, \vec{k})$ reads:

$$\begin{aligned} G(\omega, \vec{k}) &= \frac{\omega + \mu_B}{(\omega - \omega_1)(\omega - \omega_2)} = \frac{(\omega + \mu_B)}{(\omega_1 - \omega_2)}\left[\frac{1}{\omega - \omega_1} - \frac{1}{\omega - \omega_2}\right] = \\ &= \left(\frac{\omega_1 + \mu_B}{\omega_1 - \omega_2}\right)\frac{1}{(\omega - \omega_1)} - \left(\frac{\omega_2 + \mu_B}{\omega_1 - \omega_2}\right)\frac{1}{(\omega - \omega_2)} + \frac{1}{\omega_1 - \omega_2} - \frac{1}{\omega_1 - \omega_2} = \\ &= \left(\frac{\omega_1 + \mu_B}{\omega_1 - \omega_2}\right)\frac{1}{(\omega - \omega_1)} - \left(\frac{\omega_2 + \mu_B}{\omega_1 - \omega_2}\right)\frac{1}{(\omega - \omega_2)} \end{aligned} \quad (34)$$

Let us check the poles structure:

$$\begin{cases} \omega - \omega_1 = \omega + \left(\mu_B - \varepsilon_k - \dfrac{Und^2}{2}\right) + \dfrac{\varepsilon_k \mu_B}{\mu_B - \varepsilon_k - \dfrac{Und^2}{2}} \\ \omega - \omega_2 = \omega - \dfrac{\varepsilon_k \mu_B}{\mu_B - \varepsilon_k - \dfrac{Und^2}{2}} \end{cases} \quad (35)$$

But $\mu_B = 2\mu - |E_b| \approx -|E_b| \approx -U(1 - nd^2)$ and $\mu_B - \dfrac{Und^2}{2} \approx -U\left(1 - \dfrac{nd^2}{2}\right)$. Of course $\left|\mu_B - \dfrac{Und^2}{2}\right| \gg |\varepsilon_k|$. Hence:

$$\omega - \omega_1 = \omega - \varepsilon_k - U\left(1 - \frac{nd^2}{2}\right) - \frac{\varepsilon_k U(1-nd^2)}{-U\left(1-\frac{nd^2}{2}\right)} =$$

$$= \omega - \varepsilon_k - U\left(1-\frac{nd^2}{2}\right) + \frac{\varepsilon_k(1-nd^2)}{\left(1-\frac{nd^2}{2}\right)} = \omega - U\left(1-\frac{nd^2}{2}\right) + \varepsilon_k\left(-\frac{nd^2}{2}\right) = \quad (36)$$

$$= \omega - U\left(1-\frac{nd^2}{2}\right) - \frac{nd^2}{2}\varepsilon_k$$

$$\omega - \omega_2 = \omega + \frac{\varepsilon_k U(1-nd^2)}{(-U)\left(1-\frac{nd^2}{2}\right)} = \omega - \frac{\varepsilon_k(1-nd^2)}{\left(1-\frac{nd^2}{2}\right)} \approx \omega - \varepsilon_k\left(1-\frac{nd^2}{2}\right)$$

In the same time the first term in (34) yields:

$$\frac{1}{\omega-\omega_1}\frac{\omega_1+\mu_B}{\omega_1-\omega_2} = \frac{1}{\omega-\omega_1}\frac{U\left(1-\frac{nd^2}{2}\right)+\frac{nd^2}{2}\varepsilon_k+\mu_B}{U\left(1-\frac{nd^2}{2}\right)+\frac{nd^2}{2}\varepsilon_k-\varepsilon_k\left(1-\frac{nd^2}{2}\right)} \approx$$

$$\approx \frac{1}{\omega-\omega_1}\frac{U\left(1-\frac{nd^2}{2}\right)-U(1-nd^2)}{U\left(1-\frac{nd^2}{2}\right)} = \frac{1}{\omega-\omega_1}\left(1-\frac{1-nd^2}{1-\frac{nd^2}{2}}\right) \approx \frac{nd^2}{2(\omega-\omega_1)}$$

(37)

The second term in (34) reads:

$$\frac{1}{\omega-\omega_2}\frac{\omega_2+\mu_B}{\omega_1-\omega_2} \approx \frac{1}{\omega-\omega_2}\frac{\varepsilon_k\left(1-\frac{nd^2}{2}\right)-U(1-nd^2)}{U\left(1-\frac{nd^2}{2}\right)} \approx$$

$$-\frac{1}{\omega-\omega_2}\frac{U(1-nd^2)}{U\left(1-\frac{nd^2}{2}\right)} \approx -\frac{1}{\omega-\omega_2}\left(1-\frac{nd^2}{2}\right)$$

(38)

Thus

$$G(\omega,\vec{k}) \approx \left[ \frac{\frac{nd^2}{2}}{\omega - U\left(1 - \frac{nd^2}{2}\right) - \frac{nd^2}{2}\varepsilon_k} + \frac{\left(1 - \frac{nd^2}{2}\right)}{\omega - \varepsilon_k\left(1 - \frac{nd^2}{2}\right)} \right] \quad (39)$$

and we completely recover the Hubbard-I approximation [1,13]. The first pole in (39) corresponds to the Upper Hubbard band (UHB). Thus $Z_{UHB} = \frac{nd^2}{2}$. The second pole corresponds to the lower Hubbard band (LHB): $Z_{LHB} = \left(1 - \frac{nd^2}{2}\right)$. Of course, $Z_{UHB} + Z_{LHB} = 1$. We can rewrite $G(\omega,\vec{k})$ as:

$$G(\omega,\vec{k}) = \frac{Z_{LHB}}{(\omega - \varepsilon_k Z_{LHB} + io)} + \frac{Z_{UHB}}{\left(\omega - U\left(1 - \frac{nd^2}{2}\right) - Z_{UHB}\varepsilon_k + io\right)} \quad (40)$$

Note that the second iteration to the self-consistent T-matrix approximation does not change the gross features of (40). Thus the antibound state yields non-trivial corrections to Landau Fermi-liquid picture already at low electron density, but does not destroy it in 2D. The simplest Hartree-Fock contribution to the thermodynamic potential $\Omega$ from the upper Hubbard band $\Delta\Omega \sim \int \Sigma(\omega,\vec{p}) G_0(\omega,\vec{p}) \frac{d^2\vec{p}}{(2\pi)^2} \frac{d\omega}{2\pi}$ with $G_0(\omega,\vec{p})$ and $\Sigma(\omega,\vec{p})$ given by (28), (29) yields $\Delta\Omega \sim Z_{UHB} n\varepsilon_F \sim n^3$.

**Engelbrecht-Randeria mode.**

For the sake of completeness let us discuss briefly the Engelbrecht-Randeria mode[14] which also corresponds to the pole of the T-matrix for $U >> W$ and $nd^2 << 1$. According to [14] it has a spectrum for $q < 2p_F$:

$$\omega_{ER} \approx \omega_q - \exp\left\{-\frac{1}{f_0}\right\} \cdot \frac{\omega_q^2}{2\varepsilon_F}. \quad (41)$$

Note that while antibound state exists also in 3D physics, the Engelbrecht-Randeria mode is specific for 2D Hubbard model.

In (41) $\omega_q = \dfrac{q^2}{4m} - 2\varepsilon_F$ and $\exp\left\{-\dfrac{1}{f_0}\right\} = nd^2$ in agreement with (1). Note that for $q=0$:

$$\omega_{ER} = -2\varepsilon_F - 2\varepsilon_F nd^2 < 0 \qquad (42)$$

The collective character of Engelbrecht-Randeria mode is connected with the fact that in the absence of fermionic background (for $\varepsilon_F = 0$) $\omega_{ER} = 0$ in (42). Moreover $\omega_{ER} < -2\varepsilon_F$. Hence this mode lies below the bottom of the band and corresponds to the binding of two holes (Recall that the antibound state lies above the upper edge of the band).

In terms of the "bosonic" chemical potential $\mu_B$:

$$\omega_{ER} \approx \dfrac{q^2}{4m} - \mu_B, \qquad (43)$$

where in terms of $\mu \approx -\dfrac{W}{2} + \varepsilon_F$, $\mu_B = 2\mu + |E_b|$ and the binding energy $|E_b| \approx W + 2\varepsilon_F nd^2$.

**Conclusion and Acknowledgements.**

We considered the excitation spectrum of the Hubbard model at low electron density, where a small parameter (gas parameter) allows a controlled expansion. On the level of the first iteration to the self-consistent T-matrix approximation we found the contribution of the T-matrix pole corresponding to the antibound state to the self-energy $\Sigma$. As a result we got a two-pole structure of the dressed one-particle Green-function which closely resembles the Hubbard-I approximation.

It would be interesting to find the possible contribution of the Upper Hubbard band to the ground-state energy or compressibility and to build the bridge between the Galitskii-Bloom Fermi-gas expansion for the ground-state energy (or compressibility) and the Gutzwiller type of expansion for the partially filled band[15] when the electron density is increased.

For the sake of completeness we also analyzed the Engelbrecht-Randeria mode which corresponds to the pairing of two holes below the bottom of the band. According to [14] this mode, when keeping the full q-dependence for $0 \leq q < 2p_F$, gives non-analytic corrections $\sim |\omega|^{5/2}$ to the imaginary part of the self-energy $\mathrm{Im}\Sigma(\omega)$ in 2D. It also contributes to the thermodynamics at $T = 0$ in the same order in density as the contribution of the antibound state: $\Delta\Omega \sim \varepsilon_F n \cdot n d^2 \sim n^3 > 0$ - amounting to an increase of the thermodynamic potential $\Omega$ [14]. Thus the Engelbrecht-Randeria mode as well as the Hubbard-Anderson mode corresponding to the antibound state yield interesting corrections to the Landau Fermi-liquid picture in 2D already at low electron density, but do not destroy it completely in contrast to the 1D-case, where we have the Luttinger liquid state and a vanishing quasiparticle residue $Z \to 0$ for $\omega \to 0$ [16].

We acknowledge helpful discussions with P.B. Wiegman, D. Vollhardt, P. Fulde, K.I. Kugel and A.F. Barabanov.

This work was supported by RFBR grants №11-02-00798 and 11-02-00741.